\documentclass[letters,usenatbib]{mn2e}
\usepackage{graphicx}
\usepackage{multirow}
\usepackage{txfonts}

\usepackage{natbib}
\title[Resonance Condition and LFQPOs]
{Resonance Condition and Low Frequency Quasi Periodic Oscillations of the Outbursting Source H~1743-322}
\author[S. K. Chakrabarti, S. Mondal, \& D. Debnath]
       {Sandip K. Chakrabarti$^{1,2}$\thanks{E-mail: sandip@csp.res.in}, Santanu Mondal$^2$ and Dipak Debnath$^2$\\
$^1$ S. N. Bose National Center for Basic Sciences, JD-Block, Salt Lake, Kolkata, 700098, India\\
$^2$ Indian Centre For Space Physics, 43 Chalantika, Garia Station Road, Kolkata, 700084, India}

\begin{document}


\maketitle

\begin{abstract}
It has long been proposed that low frequency QPOs in stellar mass black holes or their equivalents
in super massive black holes are results of resonances between infall and cooling time scales. We explicitly 
compute these two time scales in a generic situation to show that resonances are easily achieved.
During an outburst of a transient black hole candidate (BHC), the accretion rate of the Keplerian disk as well 
as the geometry of the Comptonizing cloud change very rapidly. During some period, resonance condition
between the cooling time scale (predominantly by Comptonization) and the infall time scale
of the Comptonizing cloud is roughly satisfied. This leads to low frequency quasi-periodic oscillations (LFQPOs)
of the Compton cloud and the consequent oscillation of hard X-rays. 
In this paper, we explicitly follow the BHC H~1743-322 during its 2010 
outburst. We compute Compton cooling time and infall time on several days and show that QPOs 
take place when these two roughly agree within $\sim 50$\%, i.e., the resonance condition 
is generally satisfied. We also confirm that for the sharper LFQPOs (i.e., higher 
Q-factors) the ratio of two time scales is very close to 1.
 
\end{abstract}

\begin{keywords}
X-Rays: binaries -- Stars:individual:H~7143-322 -- Black Holes -- shock waves -- accretion -- accretion disks -- Radiation:dynamics
\end{keywords}


\section{Introduction}

Low frequency Quasi-Periodic Oscillations or LFQPOs are common features observed in X-rays emitted from 
stellar mass black holes. X-ray transient sources in our galaxy exhibit various types of QPOs with frequencies 
ranging from mHz to a few hundreds of Hz (Morgan, Remillard \& Greiner
1997; Paul et al. 1998). Since a black hole does not have hard surface, emission of oscillating hard X-rays
is very puzzling. Several models are present in the literature to explain the origin of the quasi-periodic
oscillations (QPOs).
Kato \& Fukue (1980) suggested trapped oscillations in gaseous disks around supermassive
black holes and tried to explain 100d time variabilities in these objects.  
Carroll et al. (1985) found oscillations of the disk using purely Newtonian potential 
in their 3D numerical simulation. `Diskoseismology' model by Nowak \& Wagoner (1991), 
uses acoustic oscillations of the disk arising out of dispersion relation. Molteni, Sponholz and 
Chakrabarti (1996), in the context of super massive black holes, 
mentioned that resonance caused by agreement of bremsstrahlung type cooling and infall time scale
in the post-shock region of an advection flow is the cause of oscillation of the emitted radiation. 
Titarchuk et al. (1998), identifies low frequency QPOs associated with the viscous magneto-acoustic resonance oscillation 
of the transition layer surrounding the Compton cloud.  
Stella and Vietri (1999) explains QPOs in terms of orbital precession and show that nearly correct 
frequencies can be generated by this way. 
Trudolyubov et al. (1999) explains LFQPO using perturbation inside a Keplerian disk while
Titarchuk \& Osherovich (2000) and Shirakawa \& Lai (2002) use 
global disk oscillation and oscillation of warped disk  for the purpose.
Rodriguez et al. (2002) and Tagger et al. (2004) 
propose that LFQPOs could be due to magnetohydrodynamic (MHD)
instability of magnetized accretion disks known as the accretion-ejection instability
(AEI; also see, Tagger \& Pellat 1999) which combines spiral and Rossby waves propagating in the disk.
Recently, propagating mass accretion rate fluctuations in hotter inner disk flow (Ingram \& Done, 2011),
and oscillations from a transition layer in between the disk and hot Comptonized flow
(Stiele et al., 2013) were used to explain the origin of QPOs. 
While these models can explain the frequency, explanation of
the strong Q-factor (ratio of frequency and the full width at half maximum), 
or coherency observed in some systems is more difficult. Since black holes
do not have hard surfaces, the so-called `beat-frequency' type models (Lamb et al. 1985) 
as is used for neutron stars cannot be used. Furthermore, none of these
models attempt to  explain long  duration continuous
observations and the evolutions of QPOs during the outburst phases of transient BHCs. 
Meanwhile, Chakrabarti (1989, 1990) showed that black holes can have boundary layers in that 
unique standing shocks can form due to strong centrifugal barrier (which act as hard surfaces) 
in the vicinity of the horizon where infalling energy may be dissipated and jets/outflows may be formed.
This post-shock region could oscillate if its cooling time scale 
roughly agrees with the infall time in this region (Molteni, Sponholz \& Chakrabarti, 1996; Chakrabarti \& Manickam, 2000).
The shock oscillation model (SOM) by Chakrabarti and his collaborators naturally explains LFQPOs. 
The post-shock region is known as the CENtrifugal pressure dominated BOndary Layer or CENBOL. 
Chakrabarti \& Titarchuk (1995, hereafter CT95), based on the viscous transonic flow solution (Chakrabarti, 1990)
postulated that higher-viscosity flow in equatorial region will be Keplerian and it is immersed inside a
lower-viscosity flow which also has much lower angular momentum. This so-called Two Component Advective Flow
solution (TCAF) not only can explain the spectral properties of a black hole, the LFQPOs
can also be explained just by making an extra assumption that they occur only when 
the cooling time scale due to Comptonization in CENBOL roughly matches with the infall-time scale in the post-shock flow.
Thus the same shock-wave, used for spectral formation, can also be used for LFQPOs.
Garain et al. (2014, hereafter GGC14), using Monte Carlo simulations coupled with hydrodynamic simulations 
clearly demonstrated that observed LFQPOs could indeed be the results of resonance phenomenon. The frequencies and 
Q-values are found to be in acceptable ranges.

The hypothesis that LFQPOs are due to resonance could be tested easily in outbursting sources, as the accretion rates 
of the Keplerian and low-angular momentum components evolve continuously and 
fulfillment of resonance condition may be easier. In fact, as we show below, in these sources, resonance condition 
may remain satisfied in several successive days and the QPOs evolve accordingly.
Movement of the shock is also driven primarily by Compton cooling (Mondal, Chakrabarti \& Debnath, 2015, hereafter MCD15).
This so-called Propagating Oscillatory Shock (POS) model (Chakrabarti et al., 2008; Debnath et al., 2010, 2013;
Nandi et al. 2012) also explains the evolution of the QPOs in both the rising and declining phases of outbursts
in several black hole candidates. Generally speaking, as the day progresses in the rising phase, the rate of cooling goes up 
with more supply of soft photons from the Keplerian component of flow,
shifting the shock closer to the black hole resulting in a reduction of CENBOL size and increasing the QPO frequency 
in the process (MSC96; Das et al. 2010; Mondal \& Chakrabarti, 2013). The reverse is true in the declining phase.

In the literature, there are mentions of three types (A, B and C) of LFQPOs (Casella et al., 2005). Type C LFQPOs are 
characterized in the power spectrum by a strong (up to $\sim 16$\% rms), 
narrow ($\nu/\Delta \nu \sim 7 - 12$), and variable peak (centroid frequency) and intensity 
varying by several percent in a few days at frequencies $\sim 15$ Hz, superposed on a flat-top noise
(FTN) that steepens above a frequency comparable to the QPO frequency. Type B LFQPOs are characterized by a relatively strong ($\sim 4\% $rms) and a 
narrow peak ($\nu/\Delta \nu \geq 6$). There is no evidence of FTN, although a weak red noise
is detected at very low frequencies.
Type A LFQPOs are characterized by a relatively weak (few percent rms) and a                       
broad peak ($\nu/\Delta \nu \leq 3$) around 8 Hz. A very low amplitude red noise is observed.

So far, no study has been made to quantify the cooling and infall time scales inside the CENBOL, and 
explicitly compute them for any outbursting source on a daily basis to check if they 
are close to each other. In this paper, we do this study for C type LFQPOs mentioned 
above, which increases monotonically during the rising (hard and hard-intermediate 
spectral states) phase of the outburst. It is possible that less sharper QPOs (Types B and A)
are also due to resonance phenomenon, either weakly resonating CENBOL (Type B) or the shockless centrifugal barrier (Type A). This
aspect would be treated elsewhere. In the next Section, we derive estimates of the time scales and
argue why resonance could be quite normal inside the post-shock region where Comptonization is the strongest cooling process
stellar black holes. For super massive black holes, other cooling processes could be important and the results
would change accordingly. In \S 3, we discuss the methodology. In \S 4,  we compute different time scales from observations and 
compare them on days when QPOs occur. Finally, in \S 5, we make concluding remarks.


\section{Possibility of Resonance in the post-shock region} 

We start with CT95 configuration of TCAF solution where a high viscosity Keplerian disk of accretion rate $\dot{m_d}$ is immersed in 
a low-Keplerian halo of accretion rate $\dot{m_h}$ which undergoes an axisymmetric shock transition at $X_s$ around 
a black hole of mass $M=m~M_\odot$, where $M_\odot$ is the mass of the Sun. Let the shock of height $H_s$ be of 
compression ratio (i.e., ratio of pre-shock to post-shock velocities $V_-$ and $V_+$ respectively) $R$. We consider 
stellar mass black holes so that Comptonization is the dominating cooling process. For supermassive black holes, where the
TCAF solution applies equally well (Chakrabarti, 1995), bremsstrahlung and synchrotron processes may also be important. This will be dealt with elsewhere.

The average number density of electrons inside the post-shock region (CENBOL) is:
$$
n_e= \frac{{\dot M}_d +{\dot M}_h}{4\pi V_+X_sH_s m_p} ,
\eqno{(1)}
$$
where, $m_p$ is mass of a proton. Total thermal energy inside the CENBOL (assuming geometrical shape of a cylinder) 
of electron temperature $T_e$ which can be radiated by inverse Comptonization is, 
$$
E_t= 3 \pi \kappa X_s^2 H_s T_e n_e,
\eqno{(2)}
$$
where each particle is assumed to have a thermal energy of $3/2~\kappa T_e$.  For a two temperature
flow, electrons cool faster than protons. Hence the proton temperature $T_p$ is higher by a factor of $(m_p/m_e)^{1/2}\sim 43$.
$T_p$ is obtained from energy conservation (Chakrabarti, 1989) at the shock location:
$$
\frac{1}{2} V_-^2 + n a_-^2 = \frac{1}{2} V_+^2 +n a_+^2 ,
\eqno{(3)}
$$
where, $n=3=1/(\gamma-1)$ is the polytropic constant for a relativistic flow of polytropic index $\gamma=4/3$, $a_-$ and $a_+$ 
are the adiabatic sound speeds in the pre- and post-shock flow respectively. Assuming totally ionized hydrogen gas, 
$a^2=\frac{2\gamma\kappa T_p}{m_p}$ at cooler pre-shock flow, $n a_-^2 \sim 0$, we get,
$$
T_p=\frac{m_p}{16\kappa} V_-^2 (1-\frac{1}{R^2}) .
\eqno{(4)}
$$

The cooling rate of the intercepted seed photons by CENBOL electrons is calculated as follows: If $F(X)$ 
be the flux of radiation at $r=X$ on the standard disk (Shakura \& Sunyaev, 1973), 
the integrated flux from $X=X_s$ outwards is given by,
$$
F_I= \int_{X_s}^{+\infty} F(X) 2\pi X dX = \frac{2\pi {\cal A}}{X_s},
\eqno{(5)}
$$
where, ${\cal A} = \frac{5 \times 10^9 {\dot M}_d}{m^2} (\frac {GM}{c^2})^3$. If a fraction $f_0$ is intercepted by 
the CENBOL and their net energy is enhanced by a factor of $\Gamma$ by  inverse Comptonization, we have
the cooling rate $\Lambda_c= F_I f_0 \Gamma$. Accordingly, the cooling time scale is given by,
$$
t_c = E_t/\Lambda_c .
\eqno{(6)}
$$
The infall time scale $t_i$ inside CENBOL comes from the infall time of matter slowed down
by shock compression and turbulence in the post-shock region. 
$$
t_i = X_s/V_+ .
\eqno{(7)}
$$
The ratio of $t_c$ and $t_i$ after some simplification is given by,
$$
\tau_r = \frac{t_c}{t_i} = 3.5 \times 10^{-4} \frac{1+A_r}{f_0\Gamma} (1-\frac{1}{R^2}),
\eqno{(8)}
$$
where, $A_r = {\dot M}_h/{\dot M}_d$ is the ratio of halo rate and the disk rate.
Note that terms inside $\tau_r$ are essential to incorporate physics of Comptonization: $A_r$ measures
relative abundance of electrons vis-a-vis seed photons, $f_0$ measures the degree of intercepted 
photons which are Comptonized, $\Gamma$ measures an average factor by which each intercepted seed photon 
energy is enhanced and finally $R$ is the factor by which bulk velocity of the electron cloud (CENBOL) is slowed down
by centrifugal pressure supported shocks and turbulence. Most importantly, Eq. (8) does not explicitly depend on the
mass of the black hole and for very large $A_r$ and strong shock, the result does not explicitly depend on
any shock characteristics or mass accretion rates. Thus, once resonance is set, it is likely to remains
there for wide variation of these parameters. This is precisely we see during an outburst and that justifies our usage of 
propagatory oscillating shock solution to explain the evolution of QPOs in all the outbursting sources.

In CT95 scenario, $A_r$, $f_0$, $\Gamma$ and $R$ can vary from $\sim 0, \ 0, \ 1, 
$ and $1$ in very soft states to $\sim 10^2-10^4, \ 0.01-0.05, \ 10-40$ and $\sim 4-7$ in very hard states
respectively. Accordingly $\tau_r$ can vary in the range $\sim 0-10$. In a typical hard state,
$A_r =1000$, $f_0=0.02$, $\Gamma=20$ and $R=4$, we get $\tau_r \sim  0.8$, i.e., a condition for resonance is fulfilled.

\section{Methodology of Solution} 

In the next Section, we use parameters obtained from actual TCAF fitted 
spectra (Debnath et al. 2014, 2015ab; Mondal et al. 2014a, hereafter MDC14) and compute 
cooling and infall time scales using equations given above. For infall time scale, we assume $R \sim R_c f_t$ where, 
$R_c$ is the shock compression ratio and $f_t \sim \lambda^2$, where $\lambda$ is the angular momentum of the 
flow at the shock (Chakrabarti \& Manickam, 2000; Debnath et al., 2013) which causes backflow of matter
in CENBOL slowing it down. To calculate cooling time scale ($t_{c}$), we obtain the total thermal 
energy from the electron number density and mean electron temperature of the CENBOL. 
We analyze data of 21 observational IDs from 2010 August 9 (Modiﬁed Julian Day, i.e., MJD = 55417)
to 2010 September 20 (MJD = 55459.7). We carry out data analysis using the FTOOLS
software package HeaSoft version HEADAS 6.12 and XSPEC version 12.7. For the generation of
source and background ``.pha" files and spectral fitting using the DBB, PL models and TCAF solution, we use
the same method as described in Debnath et al. (2014). The 2.5-25 keV PCA background subtracted spectra are
fitted with TCAF-based model fits file. To achieve the best fit, a Gaussian line of peak energy
around 6.5 keV (iron-line emission) is used. For the entire outburst, we keep the hydrogen
column density ($N_H$) fixed at $1.6\times10^{22}$ atoms $cm^{-2}$ for the absorption model wabs and assume
a 1.0\% systematic error (MDC14). After achieving best fit based on a reduced $\chi^{2}$
value ($\chi_{red}^2 \sim 1$), the ``err" command is used to find 90\% confidence error values
for the model fitted parameters. 

Using the model fitted parameters, namely, disk accretion rate $\dot{m_d}$, halo accretion rate $\dot{m_h}$, location
of the shock $X_s$ and compression ratio $R$ (which includes turbulent factor), we run CT95 
(suitably changed to include weaker shocks as well) code to extract electron number density 
($n_e$) and the average temperature of the CENBOL ($T_e$). We obtain the angular momentum 
($\lambda$) of the flow using MCD15, where we choose a set of energy and angular momentum 
from the parameter space available for shock formation (Chakrabarti et al., 1989). 
After cooling, shock moves inward to satisfy
the Rankine-Hugoniot condition and finds another mean location. Angular momentum is also transported
iteratively to derive the shock location in all the days when QPOs occur. 
Finally, we use Eq. (6) and (7) to obtain the time cooling and infall time scales. 

\section{Results}

In Table 1, we show the summary of our results. For a particular observation 
of H1743-322 we obtain Keplerian disk rate ($\dot{m_d}$), sub-Keplerian rate 
($\dot{m_h}$), location of the shock ($X_s$) and compression ratio (R) by TCAF fitting. 
These parameters determine the electron number density and temperature of the CENBOL. Hydrodynamics 
(angular momentum) of the flow is determined by the amount of cooling and 
Rankine-Hugoniot shock conditions. In Cols. 1 \& 4, 
we show the observation days and cooling rate $\Lambda_c$.
On the state transition day when the object went from
hard to the hard-intermediate state (day=4.00; see Debnath et al. 2013; MDC14), the optical depth of 
the CENBOL becomes maximum (=$1.1$). This is shown in Col. 5, due to the sudden rise 
in $\dot{M_+} \surd X_s$ (see, Fig. 2). In Cols. 9 \& 12, we show the Q-factor 
(=QPO frequency/FWHM) and ratio of two time scales ($\frac{t_c}{t_i}$). 
We see that cooling increases progressively till the day (=7.81, assuming the first observation to have 
taken place on the 0th day) when QPO frequency was the highest. 
On this transition day, the object went from the hard-intermediate to the soft-intermediate state.
The ratio of the time scales becomes $\sim 1$ ($t_c \sim t_i$) and satisfies the 
resonance condition almost exactly. We see that QPO frequency is also maximum. 
On day no. 14.74, amount of cooling is maximum. This is because of sudden rise in the disk 
rate. A pure soft state starts and no QPO was found. After this date, cooling rate decreases in 
the whole soft state because of the high disk rate and small size of the CENBOL.
In Col. 6, we see that the spectral slope ($\alpha$) also increases 
and continues to the soft state when $\alpha \sim 3$. 

\begin {figure}
\vskip 0.4 cm
\centering{
\includegraphics[height=7.0truecm,width=8.0truecm]{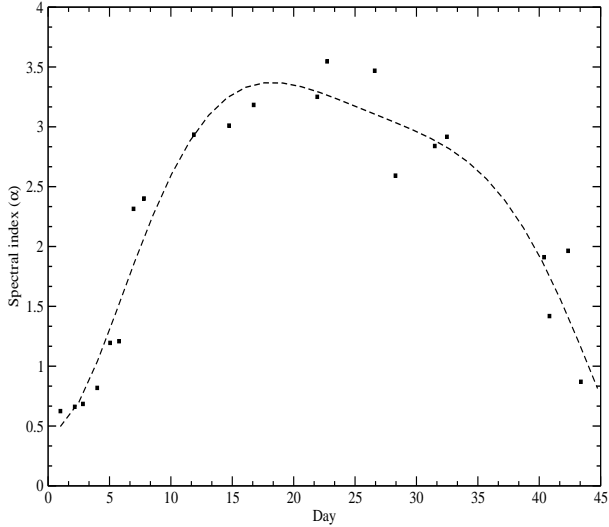}}
\caption{
Variation of energy spectral index of H~1743-322 during its 2010 outburst. 
The source started from the hard state and went to the soft state through hard- and soft-intermediate states
and then again back to the hard state. Table 1 shows the time scales and their ratio on each of the days of 
observation.}
\label{fig1}
\end {figure}

\begin {figure}
\vskip 0.4 cm
\centering{
\includegraphics[height=7.0truecm,width=8.0truecm]{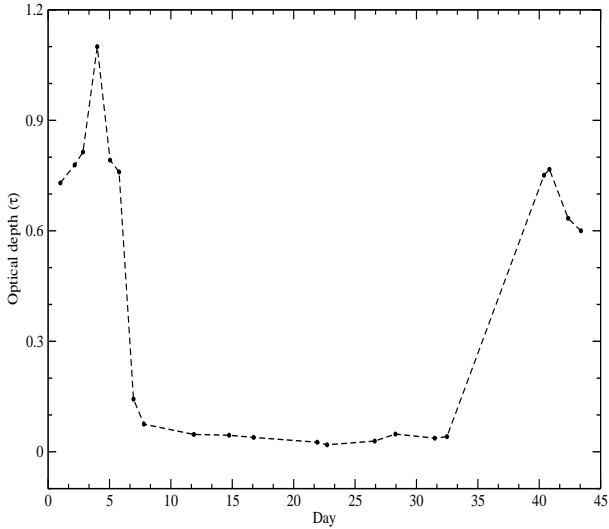}}
\caption{Daily variation of CENBOL optical depth during the 2010 outburst of H~1743-322. During the state transition 
day optical depth becomes $\sim 1$, after which it decreases. It changes in a reverse order
in the declining phase of the outburst. See the text for details.
}
\label{fig2}
\end {figure}

Note the qualitative difference in the QPOs in the rising and 
declining phases: in the rising phase, the ratio $\tau_r$ is less than unity and QPOs are seen with very high 
Q-factor. On the other hand, in the declining phase, $\tau_r$ is more than unity and Q-factor rapidly decreases. 
Most importantly, in the soft states, 
$\tau_r$ increases to a large value far away from the resonance condition, and not surprisingly,
the QPOs are also absent. 
In Fig. 1, we show the variation of spectral index for the
whole outburst. From the rising hard to declining hard state spectral index initially increases. In
the soft state it becomes more or less constant and it again decreases. 
Similarly, the variation of optical depth in Fig. 2
shows that it varies in more or less symmetric manner in both the rising and declining phases of the outburst. 
In Fig. 3, the ratio of the time scales ($\tau_r$) are plotted with day number. In the horizontal shaded
region where $0.5<\tau_r<1.5$, QPOs are expected to be observed, while in the vertical shaded
region, QPOs are actually observed during the outburst time. When QPOs are absent, the resonance condition is
clearly violated also. Thus our hypothesis that LFQPOs are seen due to resonance effect is established.
\begin {figure}
\vskip 0.4 cm
\centering{
\includegraphics[height=7.0truecm,width=8.0truecm]{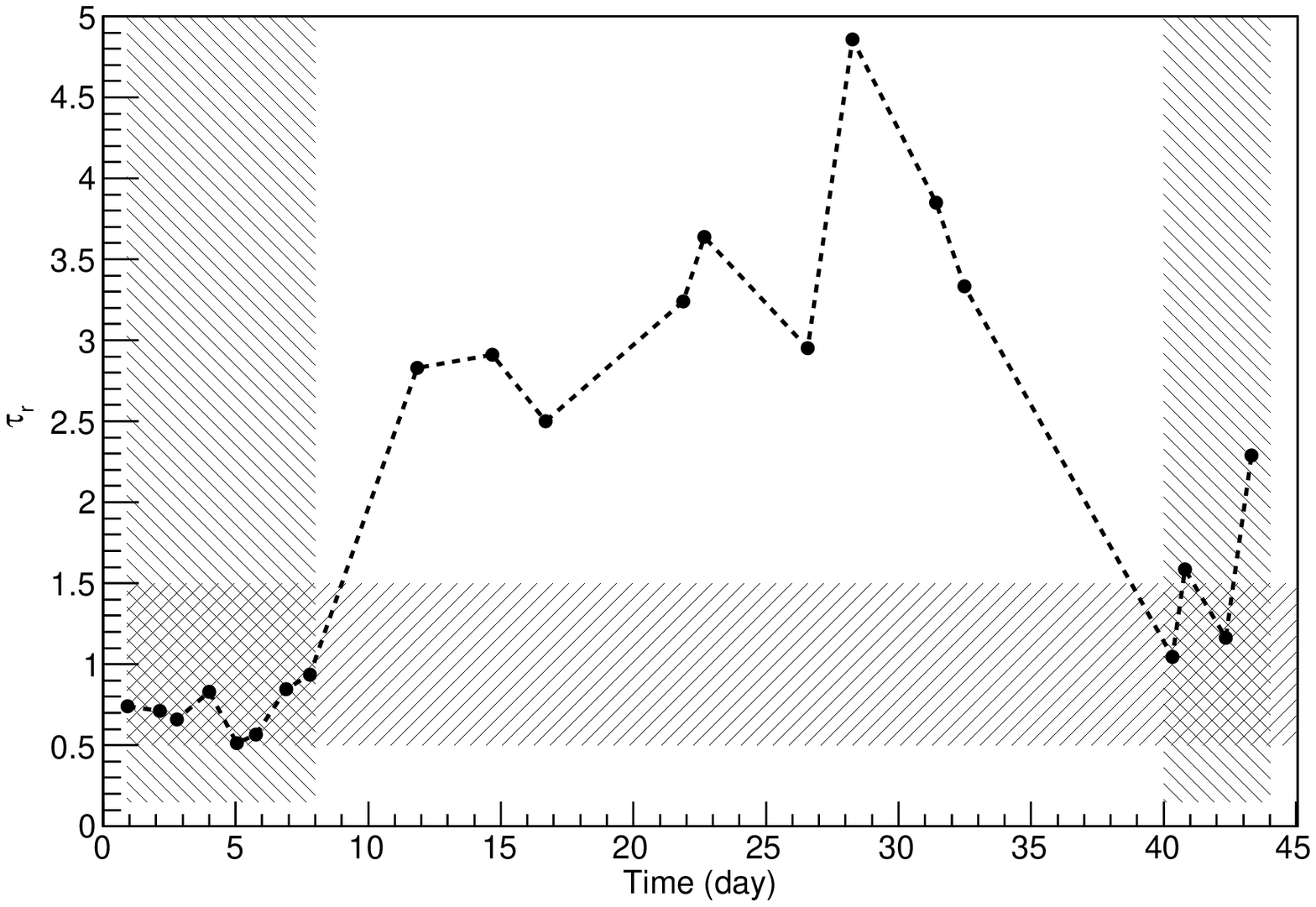}}
\caption{
The ratio $\tau_r=t_c/t_i$ is plotted against the day of the outburst. In the horizontal shaded
region where $0.5<\tau_r<1.5$, QPOs are expected to be observed, while in the vertical shaded
region, QPOs are actually observed.
}
\label{fig3}
\end {figure}

\begin{table}
\addtolength{\tabcolsep}{-4.5pt}
\centering
\vskip 0.2 cm
\centerline {Table~1: Calculated parameters for H~1743-322}
\vskip 0.2cm
\begin{tabular}{lccccccccccc}
\hline
days&${\dot m}_d$&${\dot m}_h$&$\Lambda_c$&QPOs&$\alpha$&$\tau$&$T_{e}$&$Q_{val}$&$t_{c}$&$t_{i}$&$\frac{t_c}{t_i}$ \\
  &        &    &$(10^{-3})$& (Hz)&       &      &  &     &$(s)$   &$(s)$&\\
\hline
1.00 &0.131&0.353&0.339&0.919&0.625&0.730&0.261&11.15&4.164&5.714&0.729 \\
2.17 &0.156&0.376&0.364&1.002&0.661&0.779&0.221&16.01&3.343&4.763&0.702 \\
2.84 &0.120&0.400&0.380&1.045&0.685&0.814&0.201&14.10&2.790&4.249&0.657 \\
4.00 &0.352&0.445&0.405&1.174&0.819&1.100&0.163&24.46&2.920&3.516&0.831 \\
5.05 &0.528&0.295&0.417&1.479&1.195&0.792&0.094&17.69&1.029&2.060&0.502 \\
5.78 &0.524&0.279&0.426&1.789&1.209&0.760&0.111&25.59&1.057&1.889&0.560 \\
6.96 &0.740&0.379&0.448&2.947&2.315&0.143&0.120&8.954&0.044&0.053&0.840 \\
7.81 &0.976&0.456&0.516&4.796&2.401&0.075&0.118&9.620&0.059&0.063&0.936 \\
-----&-----&-----&-----&-----&-----&-----&-----&-----&----- &-----&---- \\
11.87&3.602&0.931&0.509& -   &2.935&0.047&0.099& -   &0.340&0.120&2.820 \\
14.74&3.050&1.169&0.689& -   &3.010&0.045&0.138& -   &0.306&0.105&2.901 \\
16.74&2.850&0.477&0.406& -   &3.183&0.039&0.091& -   &0.140&0.056&2.494 \\
21.92&2.565&0.553&0.339& -   &3.251&0.026&0.098& -   &0.321&0.099&3.231 \\
22.72&2.633&0.430&0.279& -   &3.547&0.019&0.091& -   &0.338&0.093&3.634 \\
26.60&1.846&0.425&0.240& -   &3.469&0.029&0.085& -   &0.287&0.098&2.929 \\
28.29&2.098&0.803&0.269& -   &2.592&0.048&0.121& -   &0.436&0.090&4.844 \\
31.49&1.579&0.451&0.222& -   &2.840&0.037&0.112& -   &0.233&0.061&3.820 \\
32.48&1.343&0.399&0.207& -   &2.917&0.041&0.103& -   &0.199&0.060&3.317 \\
-----&-----&-----&-----&-----&-----&-----&-----&-----&-----&-----&----  \\
40.39&0.650&0.362&0.165&3.276&1.911&0.751&0.060&8.176&0.078&0.076&1.026 \\
40.83&0.718&0.345&0.154&2.569&1.419&0.767&0.086&6.937&0.138&0.087&1.648 \\
42.35&0.725&0.240&0.133&1.761&1.964&0.634&0.058&1.715&0.126&0.110&1.145 \\
43.39&0.249&0.310&0.119&1.172&0.870&0.600&0.220&1.495&0.427&0.187&2.284 \\
\hline
\end{tabular}
\leftline{In the table ${\dot m}_d$ and ${\dot m}_h$ are in Eddington unit $({\dot M}_{Edd})$.} 
\leftline{The temperature $T_e$ is in $10^{10}~K$ unit.}
\end{table}

\section {Concluding Remarks}

In this paper, we show theoretically why and how LFQPOs are formed in a stellar mass black hole environment.
Specifically we established that LFQPOs will occur when the cooling time scale due to 
Comptonization is comparable to the infall time scale (say, roughly within a factor of two). 
It is easier to verify this when the accretion rates vary widely in a short time scale. In outbursting 
sources this possibility posits itself most naturally. Furthermore, in CT95 
scenario, since the Compton cloud itself is dynamically evolving, both the cooling and the 
infall time scales evolve simultaneously in a way that the resonance condition is 
satisfied for a large number of days. Once the resonance sets in, it remains that way
for quite some time till the system evolves to extreme situations (too hard or too soft)
so that the resonance condition is grossly violated. Precisely, due to this, monotonic evolution
of the frequency of LFQPOs are observed in several transient BHCs during hard and hard-intermediate 
spectral states (Debnath et al., 2008, 2013; Nandi et al. 2012). 
Earlier in MSC96 and GGC14, this phenomena was shown through numerical simulations using bremsstrahlung 
and Compton cooling respectively. In fact, MSC94 was written with bremsstrahlung cooling with a super-massive black hole.
For the first time, we compute the ratio of time scales  on a daily basis
for a particular outbursting source, namely, H~1743-322, as exactly as possible, by first fitting the 
spectral data with TCAF solution and extracting all the disk parameters. It is already well established 
from analytical (CT95; Mondal et al. 2014b), numerical simulation (GGC14) and 
observation (Debnath et al. 2008; 2015ab; MDC14) that
the state transitions are due to the daily variations of disk and halo accretion rates.
Using the TCAF fitted parameters, we calculate the temperature ($T_e$) of the CENBOL 
as well as the electron number density ($n_e$), directly from the theoretical code (CT95). 
Turbulence which slow down accretion, requires a knowledge of angular momentum
at the shocks and this was computed using considerations presented in MCD15. 

Physically, an oscillation of the shock sets in due to the following reason: suppose at a given 
moment the shock was in a receding phase. Seen from a co-moving observer sitting on the shock, 
the flow entering upstream would
be faster than the unperturbed flow and thus the post-shock region would be hotter and cooling would be faster.
The shock will thus turn back due to reduction of post-shock pressure downstream and continue to move till
the centrifugal barrier pushes it outwards. This oscillation would thus set-in only when the cooling
and infall time scales become comparable. Too fast cooling would cause the shock to settle quickly
at a location closer to the black hole (over-damped system). Too slow cooling would be totally ignored and the 
shock would not shift at all. Thus, we see that if the disk parameters are right enough
such that $t_i$ roughly comparable with $t_c$ of the post-shock region (say within 50\%), then
the shock would oscillates radially around the mean shock location. The modulating size of the 
Post-shock region (CENBOL) would cause the intercepted photon number to also modulate around $f_0$
and thus LFQPOs are produced having a frequency inverse of the resonance time scale. In the present paper,
we concentrated on the Type C LFQPOs. However, we believe that for B type also similar resonance 
is responsible. For a large region of the parameter space, shocks are not formed, but nevertheless, 
the centrifugal barrier forms. Here a rough resonance would create LFQPOs of very low Q factors as is
seen in Type A LFQPOs, due to the absence of a sharp outer boundary (shock) of the resonating region. It is possible
that the sporadic QPOs observed in soft-intermediate states are formed due to oscillating shocks where 
the Rankine-Hugoniot conditions are not satisfied, and yet the flow has three sonic points (Ryu et al. 1997). Here,
the system is already far away from resonance condition, and thus Type C LFQPOs cannot form.

So far, we concentrated on TCAF solutions in stellar mass black holes. However, as shown in Chakrabarti (1995),
the solution is equally valid for super-massive black holes. All the theoretical considerations would remain valid,
except that possibly cooling other than Comptonization may also be important. In that case our Eq. (6) would 
require respective cooling processes. In future, we shall apply our understanding 
to other outbursting sources such as XTE~J1550-564, GRS~1915+105, MAXI~J1659-152, MAXI~J1836-194 etc.
and also to super-massive black holes. This will be dealt with elsewhere.

\noindent {\bf Acknowledgments:} 

SM acknowledges support from a post doctoral grant 
of the Ministry of Earth Science, Govt. of India. DD acknowledges supports from the project 
funds of DST sponsored Fast-track Young Scientist and ISRO sponsored RESPOND.


{}


\begin{thebibliography}{}


\bibitem[Carroll et al.(1985)]{Carroll85}Carrol, B. W., et al., 1985, ApJ, 296, 529
\bibitem[Casella et al.(2005)]{Casella05} Casella, P., Belloni, T., Stella, L., 2005, ApJ, 629, 403
\bibitem[Chakrabarti(1989)]{C89} Chakrabarti, S.K., 1989, ApJ, 347, 365
\bibitem[Chakrabarti(1990)]{C90}Chakrabarti, S.K., 1990, ``Theory of Transonic Astrophysical Flows", World Scientific (Singapore)
\bibitem[Chakrabarti(1994)]{C94} Chakrabarti, S.K., 1995, in the Proceedings of 17th Texas Symposium (1994), 
p. 546, ed. H. B\"ohringer, E. Morfill \& J. Tr\"umper (New York Academy of Sciences, New York).
\bibitem[Chakrabarti \& Titarchuk(1995)]{CT95} Chakrabarti, S.K., Titarchuk, L.G., 1995, ApJ, 455, 623 (CT95)
\bibitem[Chakrabarti \& Manickam(2000)]{CM00} Chakrabarti, S.K., Manickam, S.G., 2000, ApJ, 531, L41
\bibitem[Chakrabarti(2008)]{C08} Chakrabarti, S.K., Debnath, D., Nandi, A., Pal, P.S., 2008, A\& A, 489, L41
\bibitem[Das et al.(2010)]{Das10}Das, S., Chakrabarti, S. K., Mondal, S., 2010, MNRAS, 401, 2053
\bibitem[Debnath et al.(2008)]{DD08} Debnath, D., Chakrabarti, S.K., Nandi, A., Mandal, S., 2008, BASI, 36, 151
\bibitem[Debnath et al.(2010)]{DD10} Debnath, D., Chakrabarti, S.K., Nandi, A., 2010, A\&A, 520, 98
\bibitem[Debnath et al.(2013)]{DD13} Debnath, D., Chakrabarti, S.K., Nandi, A., 2013, AdSpR, 52, 2143 
\bibitem[Debnath, Chakrabarti \& Mondal(2014)]{DD14} Debnath, D., Chakrabarti, S.K., Mondal, S., 2014, MNRAS, 440L,121
\bibitem[Debnath, Mondal \& Chakrabarti(2015a)]{DD15a} Debnath, D., Mondal, S., Chakrabarti, S.K., 2015a, MNRAS, 447, 1984
\bibitem[Debnath, et al.(2015b)]{DD15b} Debnath, D., Molla, A. A., Chakrabarti, S.K., Mondal, S., 2015b, ApJ (in press)
\bibitem[Garain, Ghosh \& Chakrabarti(2014)]{GGC14} Garain, S., Ghosh, H., Chakrabarti, S.K., 2014, MNRAS, 437, 1329
\bibitem[Ingram \& Done(2011)]{Ingram11}Ingram, A., Done, C., 2011, MNRAS, 415, 2323
\bibitem[Kato \& Fukue(1980)]{Kato80}Kato, S., Fukue, J., 1980, PASJ, 32, 377
\bibitem[Lamb et al.(1985)]{Lamb85}Lamb, F. K., Shibazaki, N., Alpar, M. A., Shaham, J., 1985, Nature, 317, 681
\bibitem[Molteni, Sponholz \& Chakrabarti(1996)]{MSC96} Molteni, D., Sponholz, H., Chakrabarti, S.K., 1996, ApJ, 457, 805 (MSC96)
\bibitem[Mondal, \& Chakrabarti(2013)]{Mondal13} Mondal, S., Chakrabarti, S.K., 2013, MNRAS, 431, 2716 
\bibitem[Mondal, Debnath \& Chakrabarti(2014a)]{Mondal14a} Mondal, S., Debnath, D., Chakrabarti, S.K., 2014a, ApJ, 786, 4 (MDC14)
\bibitem[Mondal, Chakrabarti \& Debnath(2014b)]{Mondal14b} Mondal, S., Chakrabarti, S.K., Debnath, D., 2014b, Ap\&SS, 353, 223
\bibitem[Mondal, Chakrabarti, \& Debnath(2015)]{Mondal15} Mondal, S., Chakrabarti, S.K., Debnath, D., 2015, ApJ, 798, 5 (MCD15)
\bibitem[Morgan et al.(1997)]{Morgan97}Morgan E. H., Remillard R. A., Greiner J., 1997, ApJ 482, 993
\bibitem[Nandi et al.(2012)]{Nandi12} Nandi, A., Debnath, D., Mandal, S., Chakrabarti, S.K., 2012, A\&A, 542, 56
\bibitem[Nowak \& Wagoner(1991)]{Nowak91}Nowak, M., Wagoner, R. V., 1991, ApJ, 378, 656
\bibitem[Paul et al.(1998)]{Paul98}Paul B., Agrawal P. C., Rao A. R., Vahia M. N., Yadav J. S., 1998, ApJ 492, L63
\bibitem[Rodriguez et al.(2002)]{Rodriguez02}Rodriguez, J., Varni\'ere, P., Tagger, M., \& Durouchoux, P., 2002, A\&A, 387, 487
\bibitem[Ryu, Chakrabarti \& Molteni(1997)]{RCM97} Ryu, D., Chakrabarti, S.K., \& Molteni, D., 1997, ApJ, 474, 378
\bibitem[Shakura \& Sunyaev(1973)]{SS73} Shakura, N.I., Sunyaev, R.A., 1973, A\&A, 24, 337 (SS73)
\bibitem[Stella \& Vietri(1999)]{Stella99}Stella, L., Vietri, M., 1999, NuPhS, 69, 135
\bibitem[Stiele et al.(2013)]{Stiele13}Stiele, H., Belloni, T., Kalemci, E., et al., 2013, MNRAS, 429, 2655
\bibitem[Tagger \& Pellat(1999)]{Tagger99} Tagger, M., \& Pellat, R. 1999, A\&A, 349, 1003
\bibitem[Tagger et al.(2004)]{Tagger04} Tagger, M., Varni\'ere, P., Rodriguez, J., \& Pellat, R., 2004, ApJ, 607, 410
\bibitem[Titarchuk et al.(1998)]{Tit98}Titarchuk, L., Lapidus, I., Muslimov, A., 1998, ApJ, 499, 315
\bibitem[Trudolyubov et al.(1999)]{Trudolyubov99}Trudolyubov, S., Churazov, E., \& Gilfanov, M., 1999, A\&A, 351, L15

\end{thebibliography}
\end{document}